\documentclass[aps,twocolumn,superscriptaddress,showkeys, nofootinbib]{revtex4-1}

\usepackage{amssymb}
\usepackage{amsmath}
\usepackage{amsfonts}
\usepackage{graphicx}
\usepackage{xcolor}
\usepackage{xspace}
\usepackage{ulem}
\usepackage{slashed}
\usepackage{mathtools}
\usepackage{hhline}
\usepackage{xcolor}
\usepackage{ulem}
\usepackage[unicode=true,pdfusetitle,
bookmarks=true,bookmarksnumbered=false,bookmarksopen=false,
breaklinks=false,pdfborder={0 0 1},backref=false,colorlinks=true]
{hyperref}

\hypersetup{%
	,urlcolor=blue
	,citecolor=blue
	,linkcolor=blue
}

\newcommand{\bq}{\begin{eqnarray}}
\newcommand{\nq}{\end{eqnarray}}

\newcommand{\AddrAveiro}{Departamento de F\'{\i}sica da Universidade de Aveiro,  Campus de Santiago, 3810-183 Aveiro, Portugal}
\newcommand{\AddrCIDMA}{CIDMA,  Campus de Santiago, 3810-183 Aveiro, Portugal}
\newcommand{\AddrCFisUC}{CFisUC, Rua Larga, 3004-516 Coimbra, Portugal}
\newcommand{\AddrCoimbra}{Univ Coimbra, Faculdade de Ci\^encias e Tecnologia da Universidade de Coimbra, Rua Larga, 3004-516 Coimbra, Portugal}

\begin{document}


\title{Spontaneous breaking of the Peccei-Quinn symmetry during warm inflation}

\author{Jo\~{a}o G.~Rosa} \email{jgrosa@uc.pt} \affiliation{\AddrCoimbra}\affiliation{\AddrCFisUC}
\author{Lu\'{\i}s B. Ventura} \email{lbventura@ua.pt} \affiliation{\AddrAveiro}\affiliation{\AddrCFisUC}\affiliation{\AddrCIDMA}

\date{\today}

\begin{abstract} 
We show that, for values of the axion decay constant parametrically close to the GUT scale, the Peccei-Quinn phase transition may naturally occur during warm inflation. This results from interactions between the Peccei-Quinn scalar field and the ambient thermal bath, which is sustained by the inflaton field through dissipative effects. It is therefore possible for the axion field to appear as a dynamical degree of freedom only after observable CMB scales have become super-horizon, thus avoiding the large-scale axion isocurvature perturbations that typically plague such models.  This nevertheless yields a nearly scale-invariant spectrum of axion isocurvature perturbations on small scales, with a density contrast of up to a few percent, which may have a significant impact on the formation of gravitationally-bound axion structures such as mini-clusters.
\end{abstract}


\maketitle





\section{Introduction} \label{Sec: Intro}

The QCD axion is one of the most promising dark matter candidates, appearing naturally within the Peccei-Quinn (PQ) solution to the strong CP-problem \cite{Peccei:1977hh, Preskill:1982cy, Dine:1982ah}. In canonical models, the axion appears as the pseudo Nambu-Goldstone boson associated with the spontaneous breaking of a global U(1) PQ symmetry under which Standard Model quarks, and possibly other fields, are charged. It has a naturally small mass due to non-perturbative QCD instantons, $m_a\sim \Lambda_{QCD}^2/f_a$, where the axion decay constant $f_a$ corresponds to the PQ symmetry-breaking scale. Astrophysical constraints yield $f_a\gtrsim 10^8$ GeV, such that $m_a\lesssim 0.1$ eV.  In this low-mass range, this ``invisible" axion decays into photon pairs with a lifetime exceeding the age of the Universe, making it a natural dark matter candidate.

The main cosmological mechanism for axion production is, in most scenarios, the so-called misalignment mechanism, where the axion field oscillates coherently about the minimum of its instanton-induced potential, mimicking a pressureless fluid. The dark matter abundance depends on the axion field's displacement from the minimum at the onset of oscillations just after the QCD phase transition, which is proportional to the axion decay constant. If the latter is close to the grand unification scale, $10^{15}-10^{16}$ GeV, as typical of e.g. string theory constructions and grand unified theories (GUT), this requires a fine-tuning of the initial misalignment angle to avoid an overabundance of dark matter \cite{Linde:1991km}.

Even if such a fine-tuning can be attributed to anthropic selection in the string landscape or other dynamical mechanisms, axion models with large decay constants are, at least within the simplest scenarios, in tension with observational data. It is typically assumed in such models that the PQ symmetry remains broken throughout inflation and is not restored during the reheating process. The axion field thus remains light during inflation and acquires significant fluctuations on large scales that later become cold dark matter isocurvature modes after the onset of field oscillations \cite{Lyth:1989pb,Linde:1991km}. Such modes are uncorrelated with the main adiabatic curvature perturbations resulting from inflaton fluctuations, and are now severely constrained by the accurate measurements of Cosmic Microwave Background (CMB) anisotropies made by the Planck satellite (see \cite{Marsh:2014qoa} for a recent take on the subject). Overall, scenarios with large axion decay constants seem to only be viable within low-scale inflationary models, which predict values of  the tensor-to-scalar ratio well below the reach of any CMB observations in the foreseeable future \cite{Marsh:2015xka} (see also \cite{Higaki:2014ooa,Nakayama:2015pba,Choi:2015zra,Schmitz:2018nhb} and references therein).

An interesting alternative scenario that can prevent the generation of large axion isocurvature perturbations is one where the spontaneous breaking of the PQ symmetry occurs during the inflationary phase and is not later restored in the radiation era. This was first proposed by Linde  \cite{Linde:1991km} in a model where the complex scalar field $S$ that breaks the global U(1) PQ symmetry is coupled to the inflaton field, $\phi$, with a scalar potential of the form:
\begin{equation}
	V(S, \phi) = \lambda \left(|S|^2 - \frac{f_a^2}{2} \right)^2 + \nu \phi^2 |S|^2~,
\end{equation}
in addition to the inflaton potential. Note that the mass of the complex field receives a contribution from the inflaton. If $\nu > 0$, the bilinear interaction acts as a stabilizer, preventing the transition from the symmetric minimum at $|S| = 0$ to the broken minimum at $|S| = f_a \sqrt{1 - (\phi/\phi_c)^2}$ until $\phi < \phi_c = \sqrt{\lambda/\nu}f_a$. The inflaton field thus plays the role of the temperature in a thermal second order phase transition, and it may be possible for the critical field value to be reached only after observable CMB scales have become super-horizon during inflation. Since the axion, corresponding to the phase of the complex $S$ field in this scenario, only becomes an independent dynamical degree of freedom after the PQ phase transition, this then prevents the growth of the troublesome axion isocurvature perturbations on large scales. In addition, even if large axion fluctuations arise on small scales, any potentially associated topological defects \cite{Sikivie:2006ni} can still be inflated away if the PQ phase transition occurs sufficiently long before inflation ends.

In such a scenario the coupling $\nu$, besides the requirement of a positive sign, must be sufficiently small to ensure that the inflaton field remains light after the PQ phase transition, with the inflaton squared mass shifting by $\Delta m_\phi^2\sim \nu f_a^2\ll H^2$. It cannot, however, be too small or otherwise the critical inflaton field value could be reached before the last 50-60 e-folds of inflation.

Although Linde's scenario is certainly parametrically viable, we argue in this work that warm inflation \cite{Berera:1995ie} offers an alternative possibility for the PQ phase transition to occur during inflation. In warm inflation, the inflaton sustains a thermal radiation bath through dissipative effects, with a slowly evolving temperature typically just below the GUT scale. The latter may include some of the Standard Model fields (see e.g.~\cite{Levy:2020zfo}) that are charged under the PQ symmetry, thus leading to a thermal contribution to the mass of the PQ scalar, $S$. Hence, we may naturally expect a thermal second order PQ phase transition to occur during inflation for axion decay constants around the GUT scale. We will explore in detail the dynamics of such a phase transition and its impact on the spectrum of axion isocurvature perturbations.
 
This work is organised as follows. Section \ref{Sec: WI} summarizes the main features of warm inflation and presents the relevant dynamical quantities for the thermal PQ phase transition, analyzed in Section \ref{Sec: WI-Axions}. Section \ref{Sec: Conclusions} summarizes our main results and conclusions, as well as prospects for future developments.

\section{Warm inflation} \label{Sec: WI}

Warm inflation accounts for interactions between the inflaton scalar field $\phi$ and other degrees of freedom, in a nearly-thermal radiation bath, \textit{during} inflation \cite{Berera:1995wh, Berera:1995ie, Berera:1999ws, Berera:2008ar, BasteroGil:2009ec}. The cosmological evolution of the inflaton-radiation system is thus given by two equations for the inflaton and radiation fluids, alongside the Friedmann equation determining the Hubble expansion rate:
\begin{gather}\label{EoMTime-init}
	\ddot\phi + 3 H \dot\phi + \Upsilon \dot\phi + \partial_\phi V(\phi) = 0~, \\
	\dot\rho_r+4H\rho_r = \Upsilon \dot\phi^2~, \\
	H^2 = \frac{\rho_\phi + \rho_r}{3M_p^2}~, \label{EoMTime-end}
\end{gather}
where $\rho_\phi =\dot\phi^2/2 + V(\phi)$ is the energy density of the inflaton, $\rho_r \equiv (\pi^2/30) g_* T^4$ is the radiation energy density, with $g_*$ denoting the number of relativistic degrees of freedom, and $M_p \equiv \sqrt{\hbar c /(8 \pi G)} \approx 2.44 \times 10^{18} \text{ GeV}$ is the reduced Planck mass. The dissipation coefficient $\Upsilon=\Upsilon(\phi, T)$ can generally be computed using standard non-equilibrium quantum field theory techniques once the interactions between the inflaton and other fields are specified \cite{Berera:2008ar, Berera:1998px, Berera:1999ws, Berera:2002sp, Moss:2006gt, Graham:2008vu, BasteroGil:2010pb, BasteroGil:2012cm}. Among the several interesting differences between conventional and warm inflation are the latter's smooth transition between the inflationary stage and a radiation-dominated era if $\Upsilon/(3H) \gtrsim 1$ close to the end of inflation, and the thermal nature of inflaton fluctuations, leading to a modified primordial curvature power spectrum \cite{Taylor:2000ze, Hall:2003zp, Graham:2009bf, Ramos:2013nsa}.

For concreteness, we will focus on the warm inflation realisation of \cite{Levy:2020zfo}, which constitutes the first implementation of warm inflation within an extension of the Standard Model. This model is based on the ``Warm Little Inflaton" scenario proposed in \cite{Bastero-Gil:2016qru}, the first to overcome the challenges in realising warm inflation within a simple quantum field theory model. In the concrete scenario of \cite{Levy:2020zfo}, the inflaton dissipates its energy through its interactions with two of the right-handed neutrinos, sustaining a thermal bath that involves the latter and the Standard Model Higgs and lepton fields. The same interactions also generate light neutrino masses through the seesaw mechanism, and the third right-handed neutrino may generate the observed baryon asymmetry through thermal leptogenesis \cite{Fukugita:1986hr} just after the end of inflation. The underlying symmetries also ensure that a stable inflaton remnant remains until the present day, possibly accounting for dark matter \cite{Rosa:2018iff} or dark energy \cite{Rosa:2019jci}.

 We note, however, that the results of the next sections should be fairly independent of this particular choice of scenario since i) most of the dynamics occurs during weak dissipation slow-roll dynamics, ii) the results below are essentially contingent on a decreasing temperature during warm inflation, which is guaranteed for $\Upsilon/(3H) < 1$ and $\epsilon_\phi \equiv (M_p^2/2) (\partial_\phi V/V)^2 < 1$. For more general information on warm inflation and a detailed analysis of the model's inflationary dynamics, we refer the reader to \cite{Levy:2020zfo} and references therein.
\begin{figure}[h!]
	\centering\includegraphics[width=\linewidth]{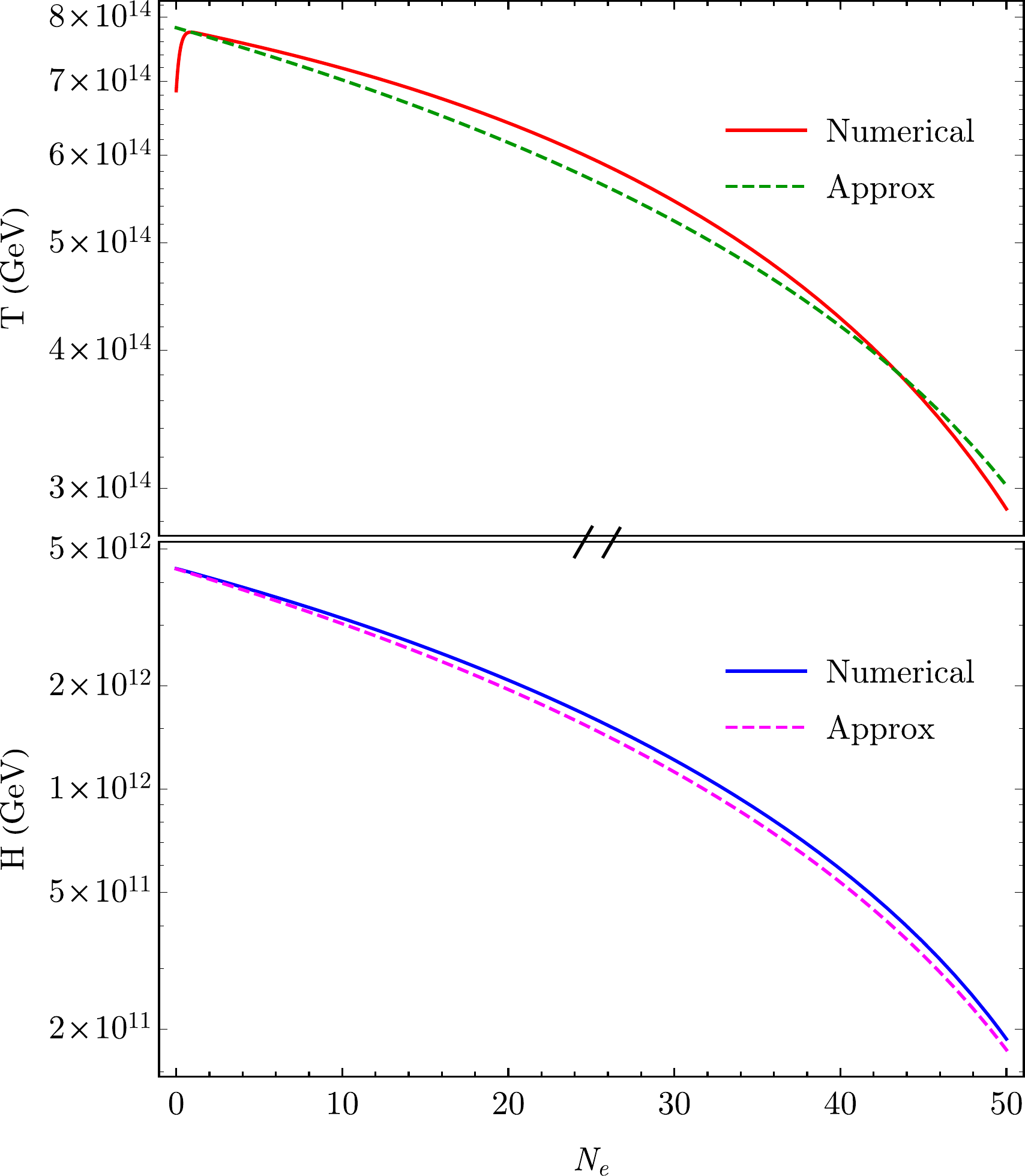} 
	\caption{Numerical evolution of the temperature $T$ and Hubble parameter $H$ (solid lines) as a function of the number of e-folds, $N_e$, in the scenario of \cite{Levy:2020zfo} for a particular choice of model parameters yielding a consistent warm inflation model. The dashed lines show the corresponding analytical approximations in Eqs. (\ref{TNe}) and (\ref{HNe}), which deviate less than 10\% from the numerical results, which is sufficiently accurate for our purposes.  \label{Fig: T and H}} 
\end{figure}

The only quantities that affect the dynamics of the PQ scalar field are the temperature of the thermal bath, $T$, and the Hubble expansion rate, $H$, during warm inflation. In Figure \ref{Fig: T and H} we illustrate the evolution of these quantities during warm inflation obtained by solving numerically the system in Eqs. (\ref{EoMTime-init}-\ref{EoMTime-end}) for a particular choice of model parameters, corresponding to the representative example shown in Figure 1 of \cite{Levy:2020zfo}. These are sufficiently well described by:
\begin{align}
	&T(N_e) = T_* \left(1 - \frac{N_e}{N_e^f}\right)^{13/20}~, \label{TNe}  \\
	&H(N_e) = H_* \left(1 - \frac{N_e}{N_e^f}\right)^{11/5} \label{HNe}~,
\end{align}
where $T_* \simeq 7.8 \times 10^{14}$ GeV, $H_* \simeq 4.4 \times 10^{12}$ GeV and $N_e^f = 65$. The quality of these approximations will be sufficient for our purposes as we aim to provide an overall description of the phase transition during warm inflation, the above expressions yielding a typical example of how the temperature and the Hubble parameter vary slowly during a period of warm inflation.




\section{Peccei-Quinn phase transition} \label{Sec: WI-Axions}

As argued above, we will assume that the complex scalar field $S$ that spontaneously breaks the Peccei-Quinn symmetry is coupled to the particles in the radiation bath during warm inflation through an effective coupling $\alpha$, which may be the result of interactions with different species in the thermal plasma. The leading correction to the PQ scalar potential is then a thermal mass correction:
\begin{equation}
	V(|S|) = \lambda \left(|S|^2 - \frac{f_a^2}{2} \right)^2 + \alpha^2 T^2 |S|^2~,
\end{equation}
where $T$ evolves during inflation according to Eq.~(\ref{TNe}) above in our working example. Here we will assume that the coupling between the PQ and inflaton fields is sufficiently small to be ignored, so that the PQ transition has a purely thermal origin, bearing in mind that there may be situations where both the inflaton coupling and the thermal correction contribute significantly to the PQ field's mass.

Decomposing the field $S$ into its radial and angular components, $\sqrt{2}S = \sigma e^{i\theta}$, the action becomes
\begin{equation}
	S = \int d^4 x \sqrt{-g} \left[\frac{1}{2}\partial_\mu \sigma \partial^\mu \sigma + \frac{\sigma^2}{2} \partial_\mu \theta \partial^\mu \theta - V(\sigma)\right]~,
\end{equation}
with
\begin{equation}
	V(\sigma) = \frac{\lambda}{4} \left(\sigma^2 - f_a^2 \right)^2 + \frac{1}{2}\alpha^2 T^2 \sigma^2~.
\end{equation}
Note that, if the expectation value of the radial field vanishes, the angular field $\theta$ has no kinetic term and is hence non-dynamical. This is a consequence of $\theta$ being ill-defined at the U(1)-symmetric minimum, $\sigma=0$. We assume that the energy density present in the PQ field is much smaller than that of the inflaton\footnote{This remains true throughout inflation when the initial amplitude of the radial field guarantees that $V(\sigma)/V(\phi) \ll 1$.}, $\rho_S \ll \rho_\phi$, so that the Hubble parameter $H$ follows from Eq. \eqref{HNe}.

The equations of motion for the radial and angular fields in a flat FRW background are, respectively,
\begin{align}
	&\ddot{\sigma} + 3 H \dot{\sigma} - \frac{\nabla^2 \sigma}{a^2}  - \sigma \dot{\theta}^2 + \partial_\sigma V(\sigma) = 0~, \\
	&\sigma^2 \left(\ddot{\theta} + 3 H \dot{\theta} - \frac{\nabla^2 \theta}{a^2} \right)  + 2 \sigma \dot{\sigma} \dot{\theta} = 0,
\end{align}
where $\dot{\sigma} = \partial_t \sigma$. Note again that, if $\sigma =0$, the angular equation of motion is trivially satisfied and only the radial field constitutes a dynamical degree of freedom.

\begin{figure}[htbp]
	\centering\includegraphics[width=\linewidth]{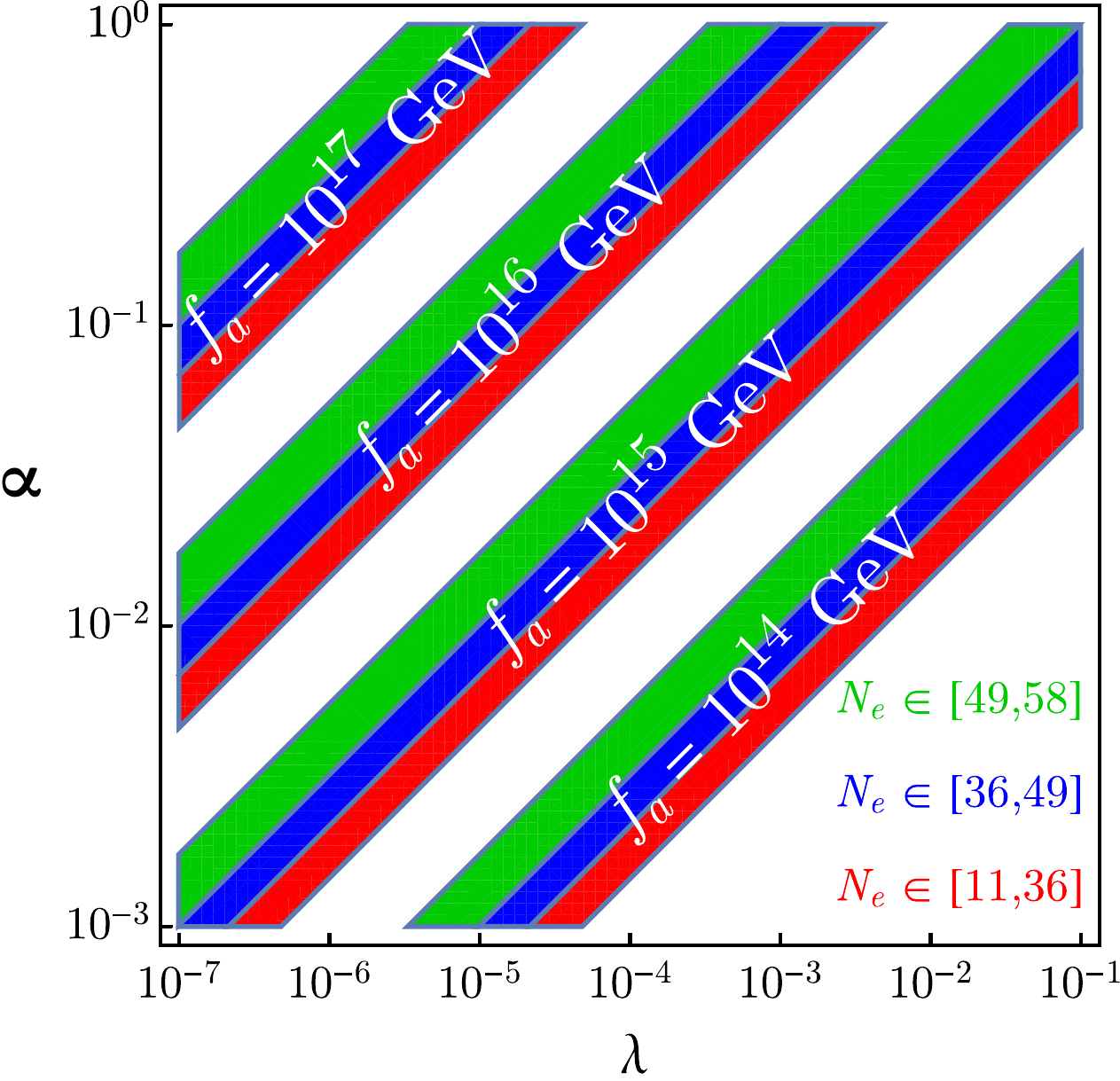} 
	\caption{Regions in the$(\lambda, \alpha)$-plane for which the PQ phase transition occurs during warm inflation, considering different values of the axion decay constant $f_a$ around the GUT scale. For each value of $f_a$, the three coloured bands given the interval in number of e-folds of inflation where the transition occurs. \label{Fig: Parameter Space}} 
\end{figure}

It is easy to see that the scalar potential has a minimum at $\sigma=0$ for temperatures above the critical temperature $T_c \equiv \sqrt{\lambda}f_a/\alpha$, corresponding to the case where the U(1) PQ symmetry is restored. Below the critical temperature, the origin becomes a maximum and the minima lie at $\sigma = f_a \sqrt{1 - (T/T_c)^2}$, which asymptotes to the axion decay constant $f_a$ at late times when $T\ll T_c$. The PQ phase transition will then occur during the last 50-60 e-folds of warm inflation if the critical temperature lies in the range $10^{14}-10^{15}$ GeV, which is typical of warm inflation models and not a special feature of our working example. This is natural for axion decay constants parametrically close to the GUT scale, and in Figure \ref{Fig: Parameter Space} we show the regions in the $(\lambda, \alpha)$-plane for which the PQ transition occurs within the last $\sim 50$ e-folds of inflation, assuming the largest observable CMB scales exit the horizon $\sim $ 60 e-folds before radiation becomes dominant.

We thus see that if neither the self-coupling of the PQ field nor its effective coupling to the particles in the thermal bath are very suppressed, a thermal PQ phase transition will naturally occur during warm inflation for large values of the axion decay constant, $f_a\sim 10^{14}-10^{16}$ GeV. We note that in this regime we always have $f_a>H/2\pi$ during inflation, but nevertheless the PQ symmetry remains restored until the temperature falls below the critical value, unlike in cold inflation scenarios \cite{Marsh:2014qoa}. 

Since during warm inflation $T\gg H$, as can be seen in the example of Figure \ref{Fig: T and H}, and we are also interested in the regime of large axion decay constant $f_a\gg H$, the PQ field effective mass will generically exceed the Hubble rate and it will oscillate about the origin with an exponentially damped amplitude for $T>T_c$. When the temperature decreases below the critical value a tachyonic instability develops driving the radial field towards the broken-symmetry minimum, about which it oscillates thereafter. It is during the instability phase when the field develops a non-trivial expectation value that the axion field becomes a dynamical degree of freedom. This simple picture is, however, incomplete, since fluctuation-dissipation effects are crucial in determining the field value at the onset of the instability, as we will now analyse in detail.


\subsection{Fluctuation-dissipation effects}

The coupling between the complex field and the radiation bath generates, in addition to the thermal mass correction, dissipative and thermal noise terms in the radial equation of motion, much like the analogous effects giving rise to warm inflation itself. The former causes the transfer of energy from the radial field to the thermal bath, while the latter prevents the amplitude of the radial field from decreasing arbitrarily as a consequence of the Hubble friction. The thermal fluctuations acting on the radial field drive its evolution, much like the stochastic collisions of water molecules with a large static pollen particle generate Brownian motion.

Since the energy density of the PQ field is much smaller than that of the inflaton field, $\rho_S \ll \rho_\phi$, we may generically neglect the radiation produced by the PQ field through dissipative effects, with the temperature of the thermal bath being essentially controlled by the inflaton field as in Eq.~(\ref{TNe}).

Adding the fluctuation-dissipation terms to the equation of motion yields an effective Langevin equation for the radial PQ field, given by (see e.g.~\cite{Berera:2008ar}):
\begin{align}
	&\ddot{\sigma} + (3 H + \Gamma) \dot{\sigma} + \lambda \sigma \left(\sigma^2 - f_a^2 \right) + \alpha^2 T^2 \sigma= \xi, \label{CRadial Eq} 
\end{align}
where $\Gamma$ is the dissipation coefficient and $\xi$ is the stochastic thermal noise, both associated with the coupling between the PQ field and radiation bath. For simplicity, we consider only a generic Yukawa coupling between the PQ field and fermions in the radiation bath of the form:
\begin{equation}
	\mathcal{L}_{\text{Yukawa}} = - g S \sum_i\overline{\psi_i} \psi'_i~,
\end{equation}
where $\psi_i$ and $\psi_i'$ denote generic light fermions in the thermal bath. The specific PQ charges of these fermions will not be relevant to our study of the phase transition, and will be model-dependent (see e.g.~the KSVZ axion model \cite{Kim:1979if,Shifman:1979if}). In this case, we have $\alpha^2 \equiv g^2 N_F/6$, where $N_F$ is the number of fermion species. The dissipation coefficient for a scalar field oscillating about the minimum of its potential coincides with its finite-temperature decay width \cite{Graham:2008vu}, which in this case is well approximated by \cite{Bastero-Gil:2016mrl}:
\begin{equation}
	\Gamma (T) \approx \frac{3 \alpha^2}{16 \pi} \frac{m_\sigma^2}{T}~, \label{Radial Diss Coef}
\end{equation}
for momentum modes below the temperature and $m_\sigma\equiv (\partial^2_\sigma V^{1/2} \lesssim T$. Note that this is valid only while the field is oscillating about the symmetric minimum for $T>T_c$. Comparing this dissipation rate with the Hubble friction yields:
\begin{equation}
	\frac{\Gamma}{H}  \approx \frac{3 \lambda \alpha^2}{16 \pi} \left(\frac{f_a}{T}\right) \left(\frac{f_a}{H}\right) \left[\left(\frac{T}{T_c}\right)^2 - 1\right],
\end{equation}
where the field mass is evaluated at $\sigma=0$. Although $f_a \gg H$, since $T \sim T_c \sim f_a$ and $\lambda, \alpha < 1$, we find that $\Gamma\lesssim H$ in most of the relevant parameter space. 

We assume a gaussian white noise term, which is a good approximation for momentum modes below the temperature of the thermal bath (see e.g.~\cite{Ramos:2013nsa, Hiramatsu:2014uta}), with a variance given by the fluctuation-dissipation relation (see e.g.~\cite{Berera:1999ws, Berera:2008ar}):
\begin{equation}
	\left<\xi_\mathbf{k}(t_1) \xi_\mathbf{k'}(t_2) \right> = 2 \Gamma (T) T \frac{(2 \pi)^3}{a^3} \delta(\mathbf{k} + \mathbf{k}')\delta(t_1 - t_2), \label{Noise Amplitude}
\end{equation}
and $\left<\xi_k(t) \right> = 0$, where the angle brackets represent a thermal ensemble average.


\subsection{Perturbation evolution}

While Hubble expansion drives the radial PQ field towards the minimum at the origin for $T>T_c$, the thermal noise induces thermal field fluctuations with a non-zero variance that affect the subsequent tachyonic instability. We may then expand the field about the symmetric minimum to yield the equation of motion for field perturbations about the origin, in momentum space:
\begin{equation} \label{perturbations_eq}
\ddot{\sigma}_k + (3 H+\Gamma) \dot{\sigma}_k + \left( \frac{k^2}{a^2} + m_\sigma^2 \right)\sigma_k = \xi_k~.
\end{equation}
Our goal is to solve this equation and evaluate the field variance on super-horizon scales, which will yield the typical oscillation amplitude of the radial field at the onset of the phase transition.

As a second order differential equation, Eq.~(\ref{perturbations_eq}) admits two homogeneous solutions, $\sigma_{k}^{\text{hom.}}(t,k) = a_kY_{1k}(t) + b_kY_{2k}(t)$, where $a_k$ and $b_k$ are coefficients depending on initial conditions. An analytical approximation to these can be found using the WKB method. In the regime where $\dot{H}/H^2 \ll 1$ and $\dot{\Gamma}/(H \Gamma) \ll 1$, one may write:
\begin{equation}
	\sigma_k^{\text{hom.}}(t) = \exp(-(3H + \Gamma)t/2) ~ u_k(t)~, \label{Decomp}
\end{equation}
so that Eq.~\eqref{perturbations_eq} becomes
\begin{equation}
	H^2 u_k'' + \left(\frac{k^2}{a^2} + m_\sigma^2 - \frac{(3H + \Gamma)^2}{4} \right) u_k \simeq 0~,
\end{equation}
where we changed integration variable to the number of e-folds, $dN_e = H dt$, $u' = du/dN_e$. Let us denote the terms inside the brackets by $f_k(N_e)$. Since $H \ll m_\sigma$ and $H$ is a slowly-varying function,  $u_k$ admits a power-series solution:
\begin{equation}
	u_k = \exp \left(\frac{1}{H} \sum_{n = 0}^\infty H^n S_n\right)~.
\end{equation}
We may then keep only the first two terms of the series and, reverting to the time variable ($t \simeq H N_e$), the two independent homogeneous solutions are:
\begin{align}
	Y_{1k}(t) \approx  \frac{A e^{-(3H + \Gamma)t/2}}{f(t)^{1/4}}\text{ cos} \left(\int_{T_*}^t dt' f_k(t')^{1/2}\right)~, \label{WKB cos sol} \\
	Y_{2k}(t) \approx \frac{B e^{-(3H + \Gamma)t/2}}{f(t)^{1/4}} \text{ sin} \left(\int_{T_*}^t dt' f_k(t')^{1/2}\right)~, \label{WKB sin sol}
\end{align}
whose amplitude is also exponentially suppressed. The inhomogeneous solution resulting from the noise term $\xi$ can be determined using the Green's function method:
\begin{equation}
	\sigma_k^{\text{inh.}}(t,k) = \int_0^t d\overline{t} ~ G_k(\overline{t},t) \xi_k(\overline{t})~, \label{Inhom Sol}
\end{equation}
where $G_k(\overline{t},t)$ is the kernel function:
\begin{equation}
	G_k(\overline{t},t) = \left[\frac{Y_{1k}(\overline{t})Y_{2k}(t) - Y_{2k}(\overline{t})Y_{1k}(t)}{W(Y_{1k}(\overline{t}),Y_{2k}(\overline{t}))}\right]~, \label{Kernel}
\end{equation}
which vanishes for $\overline{t} = t$. $W(Y_{1k}(t),Y_{2k}(t))$ is the determinant of the Wronskian matrix:
\begin{equation}
	W(Y_{1k}(t),Y_{2k}(t)) = Y_{1k}(t)\partial_t Y_{2k}(t) - Y_{2k}(t)\partial_t Y_{1k}(t)~.
\end{equation}

The structure of \eqref{Kernel} guarantees independence from initial conditions, and so one can compute $Y_{1k}(t)$ and $Y_{2k}(t)$ numerically using random initial conditions and then plug these into Eq.~\eqref{Inhom Sol} to find the inhomogeneous solution. Alternatively, one can find approximate analytical solutions using the WKB approximation above.

Note that, by definition, the inhomogeneous solution does not decay with expansion \cite{Bastero-Gil:2016mrl}. This can be seen by using the decomposition \eqref{Decomp} in Eqs.\eqref{Inhom Sol}-\eqref{Kernel}:
\begin{equation}
	\sigma_k^{\text{inh.}}(t,k) =  e^{-(3H + \Gamma)t/2} \int_0^t d\overline{t} ~  e^{(3H + \Gamma)\overline{t}/2} F_k(\overline{t},t) \xi_k(\overline{t})
\end{equation}
where the exponential factors cancel after the integration and $F_k(\overline{t},t)$ is defined as
\begin{equation}
	F_k(\overline{t},t) = \left[\frac{u_{1k}(\overline{t})u_{2k}(t) - u_{2k}(\overline{t})u_{1k}(t)}{W(u_{1k}(\overline{t}),u_{2k}(\overline{t}))}\right].
\end{equation}
As a result, $\sigma_k^{\text{inh.}}$ becomes the dominant contribution to the perturbations for $(3H+\Gamma) t \gg 1$. The combined effect of the super-horizon perturbations will then determine the evolution of the average field $\overline{\sigma}$. For this reason we shall denote the inhomogeneous solution simply as $\sigma_k(t)$.


\subsection{Thermal field variance} \label{SSec: Background field var}

Even if the radial field has a non-vanishing expectation value at the start of inflation, it will quickly be driven to the origin by the exponentially fast expansion.  There is, therefore, a point where the contribution of the super-horizon modes to the average radial field becomes dominant and the evolution thereafter is described by their combined behavior, which is captured in the two-point correlation function and field variance. With \eqref{Inhom Sol}, we can determine the former:
\begin{equation}
	\begin{aligned}
			\left<\sigma_{k}(t) \sigma_{k'}(t') \right> = \int_0^t d\overline{t} \int_0^{t'} d\tilde{t} ~ &G_k(\overline{t},t) G^{k'}(\tilde{t},t') \\
			&\times \left<\xi_k(\overline{t})\xi_{k'}(\tilde{t})\right>~.
	\end{aligned}
\end{equation}
Using the fluctuation-dissipation relation \eqref{Noise Amplitude} and selecting $t' = t$ we obtain the equal-time correlator:
\begin{equation}
	\begin{aligned}
		\left<\sigma_{k}(t) \sigma_{k'}(t) \right> = (2\pi)^3 \delta(\mathbf{k} + \mathbf{k}')  &\int_0^t d\overline{t} ~ G_k(\overline{t},t)^2 \\
		&\times \frac{2 \Gamma (T) T}{a^3}~.
	\end{aligned}
\end{equation}
This can be determined analytically by using Eqs. \eqref{WKB cos sol}-\eqref{WKB sin sol}. The kernel function is:
\begin{equation}
	G_k(\overline{t},t) \approx \frac{e^{-(3H + \Gamma)(t-\overline{t})/2}}{(f_k(t)f_k(\overline{t}))^{1/4}} \sin\left(\int_{\overline{t}}^t dt' f_k(t')^{1/2}\right),
\end{equation}
so that the two-point function becomes:
\begin{equation}
	\begin{aligned}
	\left<\sigma_{k}(t) \sigma_{k'}(t) \right> \approx \delta(k) \frac{e^{-\Gamma t}}{f_k(t)^{1/2}} \int_0^t d\overline{t} ~ \frac{2 \Gamma(\overline{t}) T(\overline{t})}{f_k(\overline{t})^{1/2}}  \times \\
	e^{\Gamma \overline{t}} \sin^2 \left(\int_{\overline{t}}^t dt' f_k(t')^{1/2}\right)~,
	\end{aligned}
\end{equation}
where $\delta(k)$ is defined as:
\begin{equation}
	\delta(k) = \frac{(2\pi)^3}{a(t)^3} \delta(\mathbf{k} + \mathbf{k}')~.
\end{equation}
Since we are interested in the contribution of super-horizon modes, $k < a H$, and $H \ll m_\sigma$, $f_k(t)$ can be approximated as:
\begin{equation} \label{f_approx}
	f_k(t) \equiv \frac{k^2}{a^2} + m_\sigma^2 - \frac{(3H + \Gamma)^2}{4}\simeq m_\sigma^2~, 
\end{equation}
such that:
\begin{equation}
	\begin{aligned}
		\left<\sigma_{k}(t) \sigma_{k'}(t) \right> \approx  \frac{3 \alpha^2}{16 \pi}\delta(k) \frac{e^{-\Gamma t}}{m_\sigma(t)} \int_0^t d\overline{t} ~ m_\sigma(\overline{t}) e^{\Gamma \overline{t}}  \times \\
		\left(1- \cos\left(2\int_{\overline{t}}^t dt' m_\sigma(t')\right)\right)~,
	\end{aligned}
\end{equation}
where we used \eqref{Radial Diss Coef} and $\sin^2(x) = (1-\cos(2x))/2$. Noting again that $H \ll m_\sigma$, it follows that the argument of the trigonometric function in the integral is rapidly oscillating, so that this contribution has a negligible average value. In the limit where $\dot{m}_\sigma/(H m_\sigma) \ll 1$,
\begin{equation}
	\begin{aligned}
		\left<\sigma_{k}(t) \sigma_{k'}(t) \right> \approx  \frac{(2\pi)^3}{a(t)^3} \delta(\mathbf{k} + \mathbf{k}') \frac{T}{m_\sigma(t)^2} 
		\left(1 - e^{-\Gamma t}\right)~, \label{Two Point Fun}
	\end{aligned}
\end{equation}
where we used \eqref{Radial Diss Coef}. This has mass dimension $-4$, as expected.

We can now compute the variance of the field generated by the contribution of all super-horizon modes. 
Returning to position space and integrating over $k$:
\begin{equation}
	\begin{aligned}
	\left<\sigma_c^2\right> = \int {d^3 k\ d^3 k'\over (2\pi)^6} \left<\sigma_{k}(t) \sigma_{k'}(t) \right>  e^{i (\mathbf{k} + \mathbf{k}')\cdot \mathbf{x}},	\label{Var Form}
	\end{aligned}
\end{equation}
and using the result \eqref{Two Point Fun} we then obtain:
\begin{equation}
	\begin{aligned}
		\left<\sigma_c^2\right> \approx \frac{\left(1 - e^{-\Gamma t}\right)}{(2\pi)^3 a(t)^3} \frac{T}{m_\sigma(t)^2} \int d^3 k~. \label{Var App Form}
	\end{aligned}
\end{equation}
The comoving momentum integral in \eqref{Var App Form} is given by:
\begin{equation}
	\begin{aligned}
		\int d^3 k 
		&=  \frac{4\pi}{3} (a(t) H)^3 \left(1 - \left(\frac{a_0}{a(t)}\right)^3 \right)~,
	\end{aligned}
\end{equation}
so that the variance \eqref{Var App Form} becomes: 
\begin{equation}
	 \left<\sigma_c^2\right> \approx \frac{1}{6 \pi^2} \frac{T H^3}{m_\sigma^2}  \left(1 - e^{-\Gamma t}\right)~, \label{Var}
\end{equation}
where we have neglected $(a_0/a)^3$, valid for $Ht \gg 1$, i.e.~after only a few e-folds of inflation.

\begin{figure}[h!]
	\centering\includegraphics[width=\linewidth]{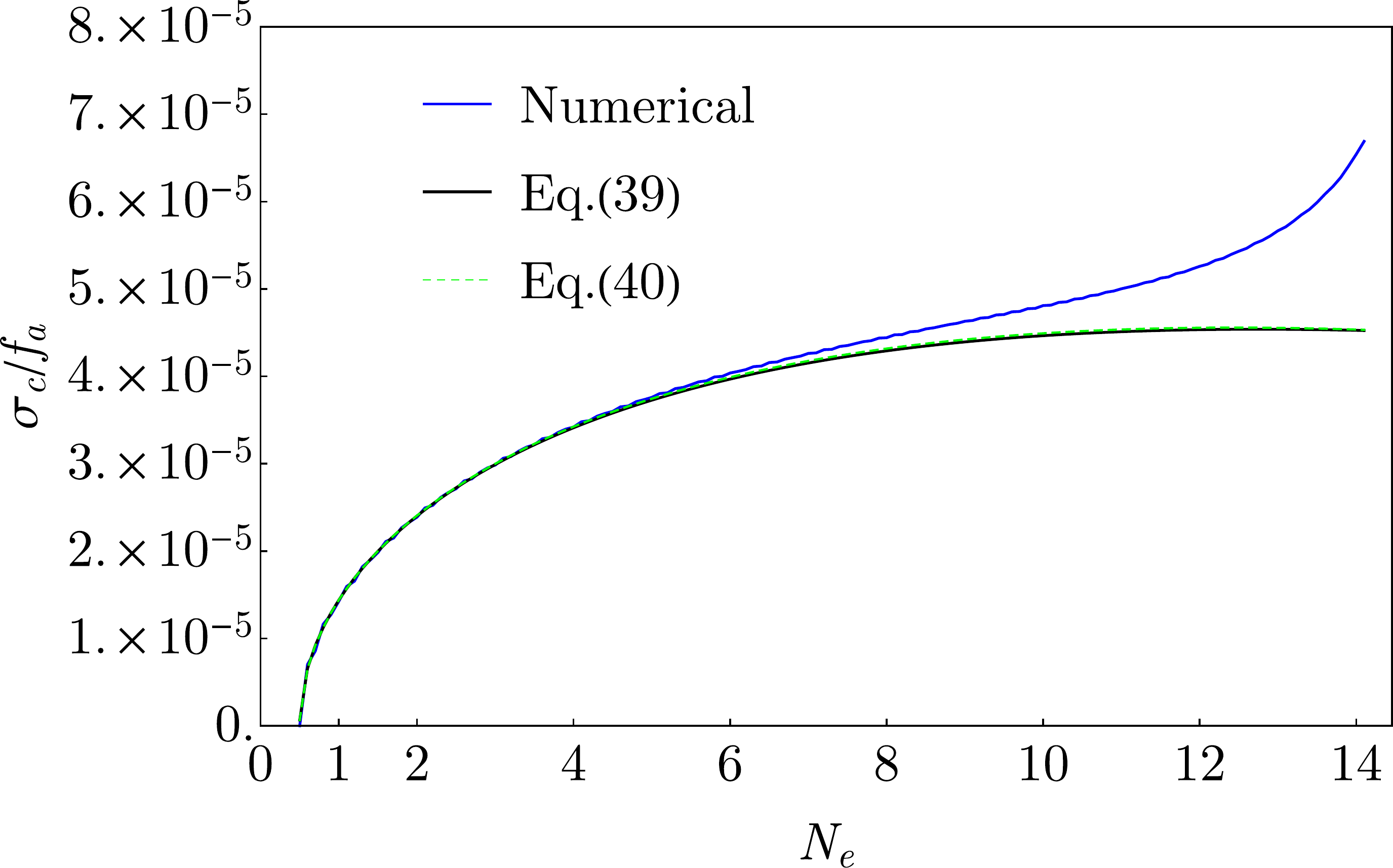} 
	\caption{Comparison between the numerical result for the square root of the variance \eqref{Var Form} determined with the numerical (blue) and the analytical approximations \eqref{Var} (black) and \eqref{Growing Var} (dashed green). The difference between the numerical and analytical results increases as $T \rightarrow T_c$ and $\dot{m}_\sigma/(H m_\sigma) \approx 1$. The microphysical parameters are $f_a = 1.2 \times 10^{15}$ GeV, self-coupling $\lambda = 0.01$ and effective coupling to the thermal bath $\alpha = 0.2$.}\label{fig_variance}
\end{figure}

For $\Gamma t \ll 1$, this can be approximated as:
\begin{equation}
	\left<\sigma_c^2\right> \simeq \frac{\alpha^2}{32\pi^3} H^3 t \simeq  \frac{\alpha^2}{32\pi^3} H^2  N_e~, \label{Growing Var}
\end{equation}
where we have used \eqref{Radial Diss Coef}.  The growing variance results from the fact that the number of super-horizon modes is increasing while their individual amplitude is not yet suppressed. This is very similar to the variance of a light quantum field in a de Sitter stage \cite{Linde:2005ht}, although fluctuations are thermal in nature.

For $\Gamma t  \gtrsim 1$, the variance can be approximated by:
\begin{equation}
	\left<\sigma_c^2\right> \simeq \frac{1}{6\pi^2} \frac{T H^3}{m_\sigma^2}~, \label{Const Var}
\end{equation}
becoming approximately constant because the growing number of super-horizon modes cannot overcome the decaying amplitude of each mode. This signals that the PQ field reaches thermal equilibrium with the ambient heat bath upon decaying significantly, as displayed by the temperature dependence and the absence of an explicit time dependence (\textit{i.e.}, the process became memoryless). However, this is only attained  for $(\Gamma/H) N_e \gg 1$, which may require a substantial number of e-folds of inflation before the phase transition as typically $\Gamma/H \ll 1$.

The typical value of the field amplitude prior to the tachyonic instability is therefore given by the square root of the variance \eqref{Var}:
\begin{equation}
	\sqrt{\left<\sigma_c^2\right>} \approx \frac{1}{\sqrt{6}\pi} H \left(\frac{T}{H}\right)^{1/2} \frac{H}{m_\sigma} \left(1 - e^{-\Gamma t}\right)^{1/2}~. \label{Typical val}
\end{equation}
We note that, although this approximate expression diverges at the critical temperature where $m_\sigma(\sigma=0)=0$, this is only due to the approximation used in Eq.~\eqref{f_approx}, while the numerically evaluated value is finite, as can be seen in Figure \ref{fig_variance}. 

Super-horizon thermal field fluctuations thus play an important role in the dynamics of the phase transition, since the PQ  never becomes too localised at the symmetric minimum, and as we will see this will determine the duration of the subsequent period of tachyonic instability, i.e.~how long the field takes to settle at the symmetry-breaking minimum.


\subsection{Tachyonic instability}

Once the minimum of the scalar potential becomes non-trivial for $T<T_c$, the dissipation coefficient is no longer given by its finite-temperature decay width, which vanishes at $T_c$ and becomes ill-defined in the tachyonic phase. However, in the latter period the field dynamics will mainly be driven by the negative sign of the field's effective squared mass, so we do not expect fluctuation-dissipation effects to play an important role.

 We will, instead, take the super-horizon RMS field amplitude computed above at the onset of the phase transition as the initial condition for the evolution of the homogeneous field component, discarding any subsequent fluctuation-dissipation effects. We are interested in determining, in particular, the point at which the field starts oscillating about the non-trivial minimum, so that the PQ symmetry is effectively broken and the angular field $\theta$ corresponding to the axion becomes an independent dynamical degree of freedom.

\begin{figure}[h!]
	\centering\includegraphics[width=\linewidth]{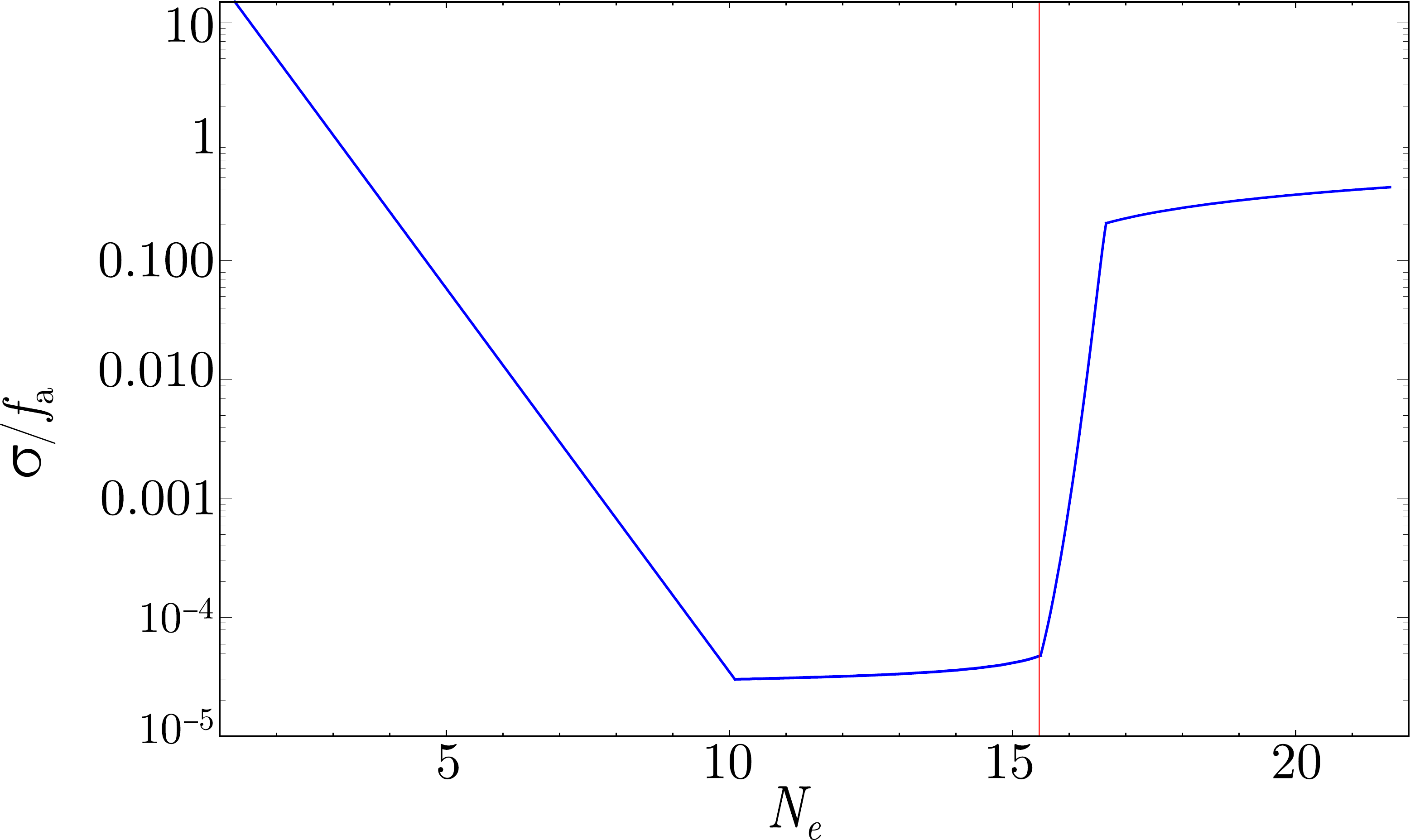} 
	\caption{Numerical evolution of the radial Peccei-Quinn field's oscillation amplitude during warm inflation in a representative example with  $f_a = 1.2 \times 10^{15}$ GeV, self-coupling $\lambda = 0.01$ and effective coupling to the thermal bath $\alpha = 0.2$. The initial amplitude of the field is $\simeq 10 f_a$, such that $V(\sigma_i)/V(\phi) \sim 10^{-3}$. The field first oscillates about the origin with an exponentially decreasing amplitude, but its amplitude cannot decrease arbitrarily due to its thermal fluctuations, which dominate the amplitude from around $N_e\simeq 10$ e-folds in this example until the critical temperature is reached around $N_e\simeq 15$ e-folds (vertical red line). At this point the tachyonic instability drives the field away from the origin and towards the temperature-dependent symmetry-breaking minimum, about which the field begins oscillating at $N_e^a\simeq 17$ e-folds. \label{Fig: radial dynamics}}
\end{figure}

The relevant equation of motion for the homogeneous field component, including all super-horizon modes, is then given by:
\begin{equation}
	\ddot{\sigma} + 3 H\dot{\sigma} + \lambda \sigma^3 + \lambda f_a^2 \left( \frac{T^2}{T_c^2} - 1\right) \sigma = 0~. \label{CRadial Eq ReWrit}
\end{equation}
We noted above that, for $\sigma < (T/T_c) f_a$, the thermal mass contribution dominates. As \eqref{Typical val} satisfies this condition, the quartic term in the potential remains negligible while the field oscillates about the origin due to the quadratic term. For $T <T_c$, a tachyonic instability develops and the field amplitude $\sigma$ grows exponentially with time. Eventually, $\sigma$ becomes large enough for the quartic term to shut down the instability: the radial field executes damped oscillations about $\sigma = f_a \sqrt{1 - (T/T_c)^2}$ thereafter. This is illustrated in the numerical example in Figure~\ref{Fig: radial dynamics}.

It is possible to obtain an approximate analytical solution during the tachyonic regime by carrying the following procedure. Firstly, we neglect the quartic contribution to the potential. Secondly, taking into account that $\dot{H}/H^2 \ll 1$, we may change variables via $\sigma = \exp(-3Ht/2) u$:
\begin{equation}
	\ddot{u} + \left( \lambda f_a^2 \left( \frac{T^2}{T_c^2} - 1\right)  - \frac{9}{4}H^2 \right)u \approx 0~. \label{Pert}
\end{equation}
Thirdly, we may also change integration variable to the number of e-folds, $dN_e = H dt$, so that:
\begin{equation}
	u'' + \left( \frac{\lambda f_a^2}{H^2} \left( \frac{T^2}{T_c^2} - 1\right)  - \frac{9}{4} \right)u \approx 0~. \label{Approximated Inst Eq}
\end{equation}
In fourth place, since $T$ varies slowly during warm inflation, we may expand $T^2$ around $T_c^2$ as a function of the number of e-folds:
\begin{equation}
	\frac{T^2}{T_c^2} \approx  1 - \gamma (N_e - N_e^c)~, \label{T near Tc} 
\end{equation}
where $\gamma > 0$ and $N_e^c$ denotes the e-fold at which $T/T_c = 1$.  Plugging $T^2/T_c^2$ into \eqref{Approximated Inst Eq}:
\begin{equation}
	u'' - \left( \frac{\lambda \gamma f_a^2}{H^2} (N_e - N_e^c)  + \frac{9}{4} \right)u \approx 0~.
\end{equation}
Shifting the integration variable, $x \equiv N_e - N_e^c$~,
\begin{equation}
	\frac{d^2 u}{dx^2} - \left( q x  + \frac{9}{4} \right)u \approx 0~, 
\end{equation}
which is simply an Airy equation with $q \equiv \lambda \gamma f_a^2/H^2 > 0$, with solutions:
\begin{equation}
	\begin{aligned}
		u(x) = A~\text{Ai}\left[q^{-2/3}\!\left(q x + \frac{9}{4} \right)\!\right] \\
		+B~\text{Bi}\left[q^{-2/3}\!\left(q x + \frac{9}{4} \right)\!\right]~,
	\end{aligned}
\end{equation}
where Ai$(z)$ is the Airy function of the first kind and Bi$(z)$ is the  Airy function of the second kind, respectively exponentially decreasing and exponentially increasing for $x > 0$. For $x < 0$, the two functions oscillate with varying frequency and decreasing amplitude as $x \rightarrow -\infty$. The constants $A$ and $B$ are fixed by the value of ($\sigma$, $\sigma'$) at $N_e \approx N_e^c$. After doing so, we can neglect the exponentially decreasing term, yielding:
%
\begin{equation}
	\sigma \approx B e^{-3 x/2} ~\text{Bi}\left[q^{-2/3} \left(q x + \frac{9}{4} \right)\right]~. \label{Approximated Inst Sol}
\end{equation} 
\vspace{0.3cm}
\begin{figure}[htbp]
	\centering\includegraphics[width=\linewidth]{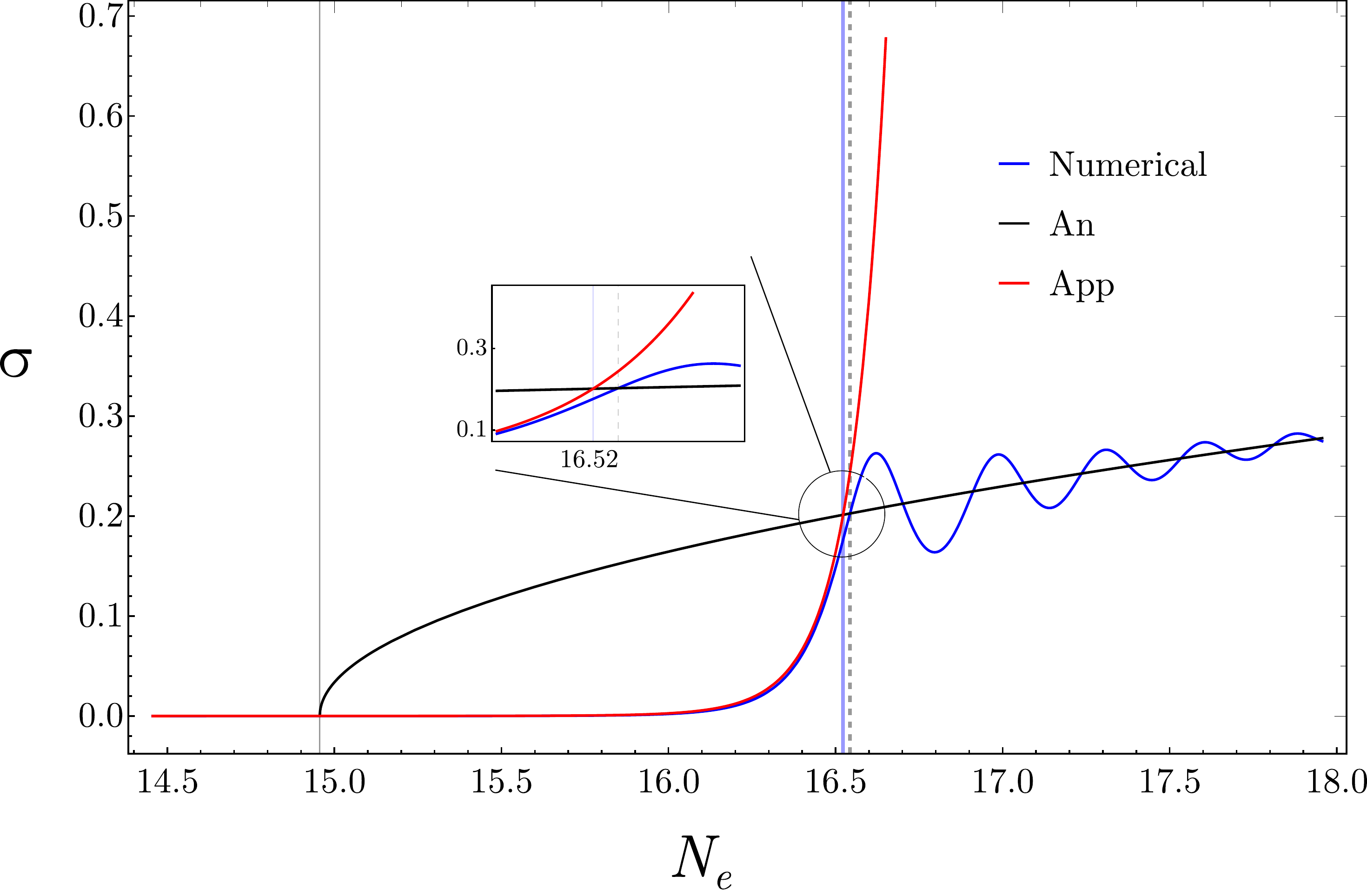} 
	\caption{Comparison between the evolution obtained numerically for the homogeneous PQ field (blue) and the analytical approximation in Eq. \eqref{Approximated Inst Sol} (red). The black curve shows the evolution of the minimum of the scalar potential, $f_a \sqrt{1 - (T/T_c)^2}$. The left-most black vertical line represents $T = T_c$, at $N_e \approx 15$. The blue (black) vertical (dashed) line represents the e-fold at which the analytical (numerical) result matches with the classical solution, Eq. \eqref{Duration of Inst}. Their relative difference is about $0.1 \%$ (see inset for zoom). The instability period lasts $x_a \simeq 1.5$ e-folds for the values of Figure \ref{Fig: radial dynamics}.\label{Fig: inst dynamics}}
\end{figure}
\vspace{0.5cm}

We can now determine the e-fold number at which the homogeneous field ``catches up" with the potential minimum $f_a \sqrt{1 - (T/T_c)^2}$, $N_e^a$ (see Figure \ref{Fig: inst dynamics}). Using \eqref{T near Tc} and \eqref{Approximated Inst Sol}:
\begin{equation}
	f_a \sqrt{\gamma x_a} \approx B e^{-3 x_a/2} ~\text{Bi}\left[q^{-2/3} \left(q x_a + \frac{9}{4} \right)\right]~,  \label{Duration of Inst}
\end{equation}
where $x_a \equiv N_e^a - N_e^c$. This equation determines the duration of the instability period $x_a$, defining how long it takes for the radial PQ field to catch up with its new minimum. We have also checked that this coincides approximately with the point at which the evolution of the radial field becomes once more adiabatic on Hubble time scales. From this we infer that $x_a$ exhibits an approximately logarithmic dependence on $B/f_a\sim \sqrt{\left<\sigma_c^2\right>}/f_a$, with a lower field variance at the critical point yielding a longer instability period, as expected. Having a sufficiently accurate determination of the duration of this instability period is essential to ascertain when symmetry breaking effectively takes place and the axion becomes an independent dynamical degree of freedom. In our numerical simulations, as shown in the example in Figure \ref{Fig: inst dynamics}, we observed generically that, at $N_e = N_e^a$, the minimum of the scalar potential lies at $\sigma/f_a=\mathcal{O}(0.1)$, with the tachyonic period lasting for $x_a\sim 1-2$ e-folds. 

It is interesting to contrast this phase transition with its equivalent during the radiation-dominated era after inflation where $a(t) \propto t^{1/2}$ and $H = 1/2t$, rather than $H \approx$ const. In this case, the solutions of \eqref{CRadial Eq ReWrit} in the unstable phase are given by Bessel functions $J_{m}(-it)$ and $Y_{m}(-it)$, where $m$ is a mass parameter that can be read from the equation of motion. Since, however, $a(t) \propto t^{1/2}$ and $N_e = \log(a/a_0) \propto \log(t)$, the elapsed time is exponential with the number of e-folds. Therefore, the unstable solution in the radiation epoch grows as an exponential of an exponential of $N_e$, rapidly catching up with the minimum of the potential at $\sigma/f_a= \sqrt{1 - (T/T_c)^2}$. As a result, $N_e^a\simeq N_e^c$ and the radial field reaches the minimum of the potential for $\sigma \ll f_a$, leading to much larger fluctuations in the phase of the field than in warm inflation as we discuss below.

After the instability, the radial field executes damped oscillations about the symmetry-breaking minimum, since its mass is much larger than the Hubble friction. As, however, $m_\sigma < T$, dissipation might still play a role on the evolution of the radial field. Describing this behavior is, however, beyond the scope of this work.


\subsection{Axion isocurvature spectrum}

The non-trivial duration of the tachyonic instability period during warm inflation has an important consequence, since the axion field is only truly ``born" after the radial field settles at the symmetry-breaking minimum. In particular, during the instability period, which as we have seen lasts for 1-2 e-folds, the Hubble-averaged value of the radial field varies in a non-adiabatic way. Since this determines the normalization of the angular kinetic term, the dynamics of $\theta$ and $\sigma$ is only effectively decoupled after $N_e^a$, when $\sigma/f_a= \sqrt{1 - (T/T_c)^2}$. 

In standard scenarios the $\theta$ field will not interact significantly with the thermal bath, but it will nevertheless exhibit quasi-de Sitter quantum fluctuations, as generic for any light scalar field during inflation \cite{Linde:1991km}:
\begin{equation}
	\delta \theta = \frac{H}{2 \pi f_a \sqrt{1 - (T/T_c)^2}}~. \label{Axion Fluctuations}
\end{equation}
We stress that this is only valid after the radial field has reached the broken symmetry minimum, so that the divergence of this expression at the critical temperature does not really yield arbitrarily large $\theta$ fluctuations. As mentioned above, we find in general that $\sqrt{1 - (T/T_c)^2}\gtrsim 0.1$, with the largest fluctuations corresponding to scales that become super-horizon around $N_e^a$. Since both the Hubble parameter and the temperature are slowly decreasing functions of the number of e-folds, $\delta \theta$ acquires a small scale dependence. This will then result in a nearly scale-invariant spectrum of isocurvature density perturbations in the dark matter, once the axion field begins oscillating about its QCD-instanton-induced potential in the radiation era.

 In order to determine whether these perturbations are significant, one needs to compute, for a given value of $f_a$\footnote{This yields a QCD axion mass $m_a \simeq 6 \times 10^{-6}/(f_a/10^{12} \text{ GeV})$ eV \cite{Linde:1991km}, with $m_a \simeq 5 \times 10^{-9}$ eV in the example above.}, the value of the initial misalignment angle $\theta_i$ for which the axion field accounts for the cosmological density of dark matter\footnote{While in \cite{Levy:2020zfo} the inflaton could account for all the dark matter, here we consider an alternative point of view where the axion accounts for most of the dark matter and the inflaton only to a sub-dominant component, which would correspond to e.g.~lower values of the inflaton mass compared to those considered in \cite{Levy:2020zfo}.}. For the QCD axion, this is given by \cite{Marsh:2015xka}:
\begin{equation}
	\Omega_A h^2 = 0.7 \left(\frac{f_a}{10^{12}\text{ GeV}}\right)^{7/6} \left(\frac{\theta_i}{\pi}\right)^2~.
\end{equation}
Using $\Omega_{\text{CDM}} h^2 = 0.13$ \cite{Aghanim:2018eyx} yields
\begin{equation}
	\theta_i \simeq 0.02 \left(\frac{f_a}{10^{15}\text{ GeV}}\right)^{-7/12}~, \label{Misalignment Angle}
\end{equation}
such that $\delta \theta_{\text{max}}/\theta_i \propto f_a^{-5/12}$ is given by the ratio of \eqref{Axion Fluctuations} and \eqref{Misalignment Angle} evaluated at $N_e = N_e^a$. For the parameters of the representative example above, $\theta_i \simeq 0.02$ and $\delta \theta_{\text{max}} /\theta_i \approx 0.06$. Since the latter value is small, the contribution of angular perturbations to the homogeneous field amplitude, $\theta_i$, can be safely neglected. 

Furthermore, it would appear in Eq. \eqref{Axion Fluctuations} that, for a fixed value of $H_*$, smaller values of $f_a$ increase the amplitude of the fluctuations in the $\theta$ field. This, however, is not true because the duration of the instability period $x_a$, which determines the point at which the radial field starts oscillating about the broken minimum, depends on the microscopic parameters ($\lambda$, $\alpha$, $f_a$). In particular, $x_a$ increases with decreasing $f_a$, compensating for the previous effect\footnote{A larger value of $f_a$ at fixed $(\lambda, \alpha)$ \eqref{Duration of Inst} leads to a larger value of $x_a$: the instability lasts longer so that $\sqrt{1-(T/T_c)^2} \rightarrow 1$ and the suppression of the denominator of \eqref{Axion Fluctuations} gradually disappears.}. Instead, for each value of $f_a$, there a particular set of parameters ($\lambda, \alpha$) that maximizes angular perturbations. We have found numerically that $\delta \rho_a/\rho_a= 2 \delta \theta/\theta  \sim \mathcal{O}(10^{-2}-10^{-1})$ for small-scale axion isocurvature perturbations.

\begin{figure}[h!]
	\centering\includegraphics[width=\linewidth]{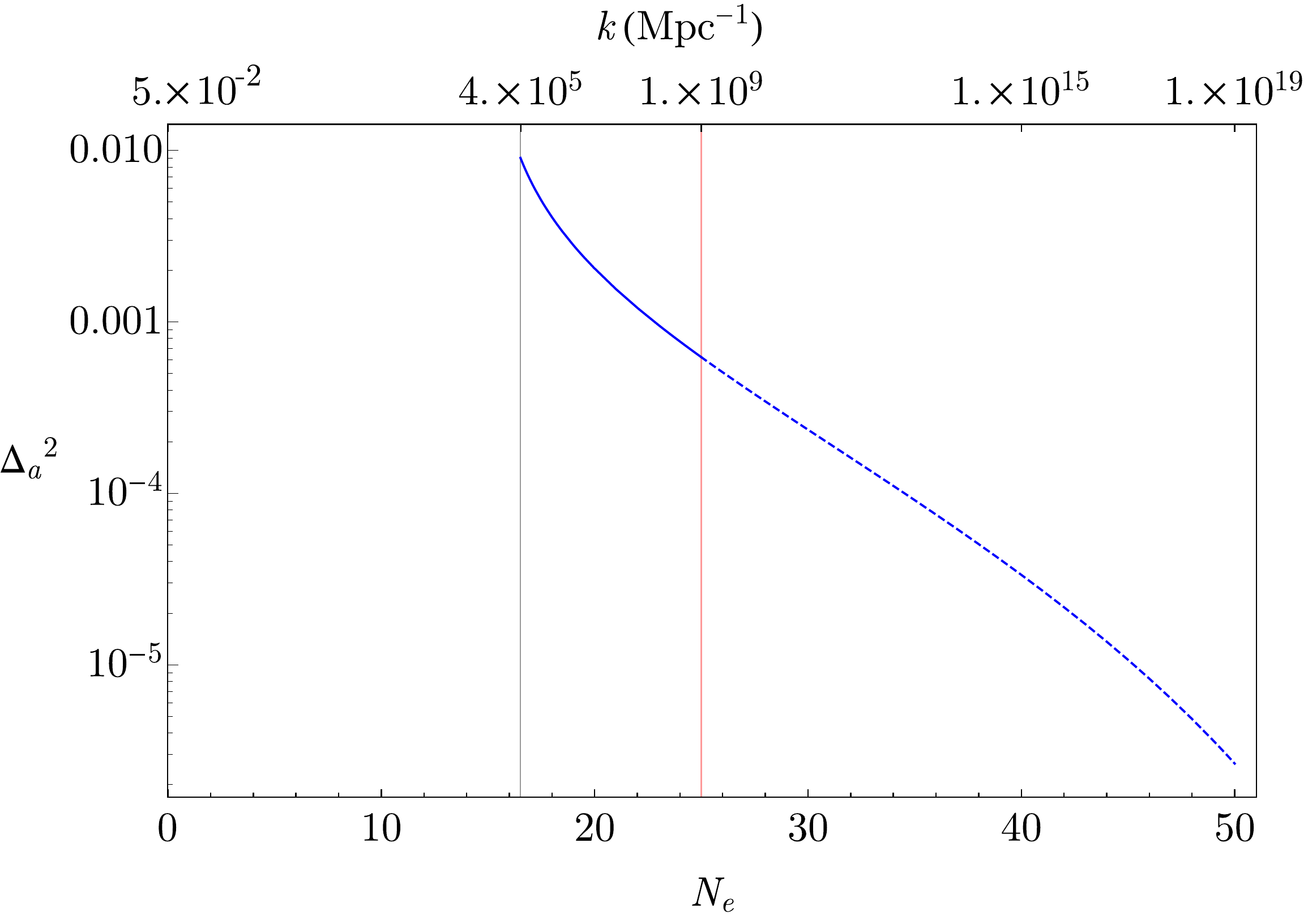} 
	\caption{Axion density fluctuation dimensionless power spectrum $\Delta_a^2 \equiv 4~(\delta \theta/\theta_i)^2$ for the example shown in Figure \ref{Fig: radial dynamics}. The vertical black and red lines correspond to the scale that leaves the horizon during inflation at $N_e^a$ (when the radial PQ field reaches the broken symmetry minimum) and the scale that re-enters the horizon at the time of the first axion oscillations in the radiation epoch, at $T \approx 1$ GeV. Perturbations on scales beyond the red line (dashed curve) become sub-horizon before the first axion oscillations and are, hence, damped away. \label{Fig: Power Spec}} 
\end{figure}

We illustrate in Figure~\ref{Fig: Power Spec} the spectrum of these perturbations as a function of the comoving wave number and the corresponding number of e-folds at which they become super-horizon, with $N_e=0$ denoting the point at which the CMB pivot scale exits the horizon during warm inflation. We note that the axion field remains massless until around the time of the QCD phase transition, when QCD instantons generate its periodic potential. Hence, fluctuations in the $\theta$ field only generate dark matter fluctuations when the temperature falls below $\sim 1 $ GeV in the radiation-dominated epoch. Fluctuations that re-enter the horizon prior to this will be smoothed out by expansion, so that our scenario only predicts significant axion isocurvature perturbations in a range of scales, the largest scale corresponding to the scale that exits the horizon during inflation when symmetry breaking effectively occurs at $N_e=N_e^a$ and the smallest scale to the one re-entering the horizon at the onset of axion field oscillations. In this range of scales, the spectrum is nearly scale-invariant, which differs significantly from the white noise spectrum associated with PQ symmetry breaking after inflation (see e.g.~\cite{Dai:2019lud}). 

\vspace{0.2cm}

This will likely affect the formation of gravitationally bound axion structures like mini-clusters and their resulting mass spectrum in a non-trivial way, particularly since the scenario of PQ symmetry breaking during warm inflation generates axion density fluctuations on super-horizon scales, as opposed to a phase transition occuring after inflation has ended. A detailed analysis of axion structure formation processes is, however, beyond the goals of this work and will be left for a dedicated future study. Nevertheless, it is clear that a PQ symmetry breaking occurring during warm inflation may have a non-trivial observational impact that may allow one to probe this hypothesis.


\section{Summary and conclusions} \label{Sec: Conclusions}

In this work we have studied a scenario where a thermal PQ phase transition, leading to the spontaneous breaking of the global U(1) symmetry proposed to solve the strong CP problem, occurs during warm inflation. This shares several similarities with the scenario proposed in \cite{Linde:1991km} in cold inflation where the PQ field is coupled to the inflaton field itself and the U(1) PQ symmetry is broken below a critical inflaton field value. In warm inflation, however, the external parameter is the (slowly evolving) temperature of the thermal bath sourced by the inflaton field through dissipative effects. In several implementations of the PQ mechanism, some of the Standard Model fields are charged under the U(1) symmetry and, provided that these are excited in the warm thermal bath (like the Higgs and lepton fields in the concrete scenario developed in \cite{Levy:2020zfo} that we have chosen as our working example), the complex field breaking the PQ symmetry naturally acquires thermal mass corrections that restore the symmetry above a critical temperature. 

Since the temperature during warm inflation typically varies in the range $10^{14}-10^{15}$ GeV and the critical temperature is parametrically close to the symmetry breaking scale, a second order thermal phase transition will naturally occur during the last 50-60 e-folds of warm inflation for values of the axion decay constant not far from the grand unification scale. This is arguably more natural than the cold inflation scenario of \cite{Linde:1991km}, where the coupling between the PQ and the inflaton fields must be tuned to yield a critical field value just below or even above the Planck scale as typical of slow-roll inflation models. Irrespective of these ``naturalness" considerations, warm inflaton certainly offers an alternative way to break the Peccei-Quinn symmetry during the last 50-60 e-folds of inflation, particularly after the large CMB scales have become super-horizon.

Our analysis took into account not only the thermal backreaction on the PQ scalar potential but also fluctuation-dissipation effects that naturally result from the coupling between the PQ field and the light degrees of freedom in the thermal bath. We have shown that, although dissipative friction is generically sub-leading compared to Hubble expansion in terms of damping the field's oscillations about the symmetric minimum, the associated thermal fluctuations prevent the field amplitude from becoming arbitrarily small. It is then the variance of the radial PQ field at the critical point that determines its subsequent evolution. We have computed this variance both numerically and through approximate analytical methods, and used it to determine the duration of the period of tachyonic instability driving the field to the symmetry-breaking minimum, showing that it typically takes 1-2 e-folds of accelerated expansion. This means that the temperature of the thermal bath when the field reaches the minimum is already slightly below the critical temperature, as opposed to what happens in a post-inflationary (second order) phase transition where the tachyonic phase occurs essentially at the critical temperature, which results in large angular fluctuations as given by Eq.~(\ref{Axion Fluctuations}) for $T\simeq T_c$.

An important consequence of the non-negligible expansion of the Universe during the tachyonic phase is the fact that the axion field, i.e.~the phase of the PQ field, only becomes a truly dynamical independent scalar field at a temperature slightly below the critical value. This limits the amplitude of its quasi-de Sitter quantum fluctuations on super-horizon scales.  Moreover, if the phase transition only occurs after the relevant CMB scales have left the horizon, axion isocurvature perturbations are only generated on small scales. The generic prediction is then a nearly scale-invariant spectrum of isocurvature perturbations in an interval of scales which (i) leave the horizon during warm inflation after the field reaches the non-trivial minimum ($N_e> N_e^a$) and (ii) reenter the horizon during the radiation era after the onset of axion oscillations ($T\lesssim 1$ GeV). In contrast, if a thermal PQ phase transition only occurs after inflation no super-horizon fluctuations are generated and the temperature remains essentially constant during the tachyonic instability period, making the initial seeds for gravitationally-bound structures in axion dark matter very different in these two scenarios, with potentially observable consequences.

The formalism that we have employed does not allow for a rigorous description of the axion isocurvature spectrum for scales that leave the horizon during the period of tachyonic instability, where radial and angular fluctuations in the complex $S$ field are not independent. This will most likely require dedicated numerical simulations of the evolution of the complex field and an accurate computation of the thermal dissipation coefficient in the tachyonic regime. Nevertheless, we expect a sharp cutoff in the axion isocurvature spectrum around the scale that leaves the horizon when the field reaches the broken symmetry minimum, suggesting that super-horizon phase fluctuations of the PQ field are quite suppressed for values of the axion decay constant around the GUT scale, as in several high-energy completions of the Standard Model. The size of these fluctuations is, in particular, below the initial phase value, $\delta\theta <\theta_i\sim 10^{-2}$ required for the axion to account for all the present dark matter abundance for such values of $f_a$. Our proposed scenario does not explain the smallness of this initial phase, which must be the result of some other pre-inflationary mechanism or simply an anthropic selection process. Our results nevertheless point towards the average PQ phase remaining close to this initial value after the phase transition has occurred during warm inflation.

We further note that it is often stated that, in the Kibble mechanism for a U(1) phase transition, the phase of the complex scalar field that breaks the symmetry cannot be correlated on super-horizon scales, which typically gives rise to topological defects. This is not, however, the case for a phase transition occurring during (warm) inflation, since accelerated expansion turns smooth Hubble patches into much larger, apparently causally disconnected regions with a large degree of homogeneity. This is, in fact, one of the primary aims of inflation. Hence, if in the initial Hubble patch that subsequently inflates the field has a phase $\theta_i$, it should maintain this phase on super-horizon scales as long as the phase remains non-dynamical. The fact that angular fluctuations below the critical temperature are suppressed compared to the assumed initial phase then means that no topological defects are expected to form in this setup, although a more accurate numerical study of the tachyonic phase is needed to support this conclusion. In any case, the absence of axion isocurvature modes on CMB scales only constrains the phase transition to occur $\gtrsim$ 10 e-folds after the largest scales become super-horizon, and the remaining period of accelerated expansion is likely more than sufficient to inflate away any topological defects that may be produced.

In summary, our analysis thus suggests that, if the Universe was warm during inflation and the PQ phase transition occurred during this period, the QCD axion is a viable dark matter candidate for large values of its decay constant close to the GUT scale, since (i) no isocurvature perturbations are generated on CMB scales, (ii) a small initial phase for the complex PQ field is stable against fluctuations; and (iii) topological defects are essentially absent in this setup.

Although our study was focused on the QCD axion and the underlying PQ model, our analysis holds for other axion-like fields. In fact, it applies to any second-order phase transition driven by a complex scalar field that is coupled to the thermal bath during warm inflation, being the first analysis of this kind\footnote{See also \cite{Laguna:1997cf, Bartrum:2014fla} for applications of fluctuation-dissipation dynamics to cosmological phase transitions.}. We hope that our results motivate further exploration of this topic.

\vspace{0.1cm}
\begin{acknowledgments}
	This work was supported by the the FCT Grant No.~IF/01597/2015. J.\,G.\,R. and L.\,B.\,V. are also supported by the CFisUC strategic project No.~UID/FIS/04564/2019. J.~G.~R. is also supported by the project ENGAGE SKA (POCI-01-0145-FEDER-022217). L.B.V was supported by the FCT grant PD/BD/140917/2019 and partially by the CIDMA project No. UID/MAT/04106/2020.
\end{acknowledgments}

\end{document}